\documentclass[prb,showpacs,preprintnumbers,amsmath,amssymb]{revtex4} 
\usepackage{graphicx}
\usepackage{bm}

\begin{document}
\title{Short Wavelength Cutoff Effects in the AC Fluctuation
Conductivity of Superconductors}
\author{
D.-N.~Peligrad and M.~Mehring}
\affiliation{2. Physikalisches
Institut, Universit\"at Stuttgart, 70550 Stuttgart, Germany}
\author{
A.~Dul\v{c}i\'{c}}
\affiliation{Department of Physics, Faculty of Science,
University of Zagreb, POB 331, 10002 Zagreb, Croatia}
\date{Submitted February 28, 2002; revised version November 26, 2002}

\thanks{present address: Philips Research Laboratories, Weisshausstrasse 2, D-52066, Aachen, Germany}

\begin{abstract}
The short wavelength cutoff has been introduced in the calculation
of $\it{ac}$ fluctuation conductivity of superconductors. It is
shown that a finite cutoff leads to a breakdown of the scaling
property in frequency and temperature. Also, it increases the
phase $\phi$ of the complex conductivity
($\tan{\phi}=\sigma_2/\sigma_1$) beyond $\pi/4$ at $T_c$. Detailed
expressions containing all essential parameters are derived for 3D
isotropic and anisotropic fluctuation conductivity. In the {\it
2D} case we obtain individual expressions for the fluctuation
conductivity for each term in the sum over discrete wavevectors
perpendicular to the film plane. A comparison of the theory to the
experimental microwave fluctuation conductivity is provided. \\
\end{abstract}

\maketitle

\section{INTRODUCTION}
\label{Sect1}

Fluctuations of the order parameter near the critical temperature
$T_c$ are much larger in high-$T_c$ superconductors than in
classical low temperature superconductors. One of the reasons lies
in the higher thermal energy $k_{\rm B}T_c$ which provides the
excitations, and the other in a very short coherence lengths which
occur in high-$T_c$ cuprate superconductors. With these
properties, the region of critical fluctuations was estimated from
the Ginzburg criterion to be of the order of 1K, or more, around
$T_c$, which renders the critical region accessible to
experimental investigations.\cite{Lobb:87}
Farther above $T_c$, one expects to
observe the transition from critical to noninteracting Gaussian
fluctuations which are the lowest order fluctuation corrections to
the mean field theory.\cite{Skocpol:75}
\par
The layered structure of high-$T_c$ superconductors requires some
theoretical sophistication. One could treat these superconductors
with various models from three-dimensional (3D) anisotropic to
coupled layers Lawrence-Doniach, or purely two-dimensional ({\it
2D}) ones. Due to the temperature variation of the coherence
lengths one could even expect a dimensional crossover in some
systems. The fluctuation conductivity is altered by dimensionality
in various models so that a detailed comparison of model
calculations and experimental data could address the
dimensionality problem.
\par
For the reasons stated above, the fluctuation conductivity in
high-$T_c$ superconductors was studied experimentally by many authors.
\cite{Ausloos:88,Hopfengartner:91,Gauzzi:95,Anlage:96,Booth:96,%
Kotzler:97,Cimberle:97,Menegotto:97,Han:98,Han:00,Silva:01,Menegotto:01}
Most of them used $\it{dc}$ resistivity measurements.
\cite{Ausloos:88,Hopfengartner:91,Gauzzi:95,Cimberle:97,Menegotto:97,%
Han:98,Han:00,Silva:01,Menegotto:01}
The reports were controversial in the conclusions about the
dimensionality of the system, and the critical exponents.
It has been shown that, in a wide temperature range above $T_c$,
the fluctuation conductivity did not follow any of the single
exponent power laws predicted by scaling and mean-field
theories.\cite{Gauzzi:95} The data in the Gaussian regime could
be fitted by an expression derived within the Ginzburg-Landau (GL)
theory with a short wavelength cutoff in the fluctuation spectrum.
Recently, Silva et al.\cite{Silva:01} have proven that the GL
approach with an appropriate choice of the cutoff parameter yields
result which is identical to that of the microscopic
Aslamazov-Larkin (AL) approach with reduced excitations of the
short wavelength fluctuations.\cite{Reggiani:91} It has been
further shown that the detailed temperature dependence of the
fluctuation conductivity was not universal, but sample dependent.
In this respect, the GL approach has practical advantage since
the cutoff parameter can be readily adjusted in fitting the
experimental data. Silva et al.\cite{Silva:01} could fit very
well the data on a number of thin films in the Gaussian region from
$T_c$+1 K to $T_c$+25 K.
\par
When critical fluctuations are studied, it becomes essential to
know accurately the value of $T_c$. However, the determination of
$T_c$ from $\it{dc}$ resistivity measurements brings about some
uncertainties. One should avoid the use of unjustifiable
definitions of $T_c$ such as e.~g.: (i) zero resistance
temperature, (ii) midpoint of the transition, (iii) maximum of the
derivative $d\rho/dT$, (iv) intersection of the tangent to the
transition curve with the temperature axis, etc. The correct value
of the critical temperature can be determined as an additional
fitting parameter in the analysis of the fluctuation conductivity.
Usually one assumes that a well defined power law holds in a given
narrow temperature range and then determine both, $T_c$ and the
critical exponent from the selected
segment.\cite{Menegotto:97,Han:00,Menegotto:01} However, the
experimental data usually show an almost continuous change of the
slope so that the uncertainty in the determination of $T_c$ is an
unsolved problem. Besides, the effects of the cutoff have been
neglected in the analysis of the data close to $T_c$. Even though
the values of the fluctuation conductivity near $T_c$ are not much
affected by the introduction of the cutoff, the slopes can be
considerably changed,\cite{Hopfengartner:91} and the analysis may
become uncertain.
\par
A number of microwave studies have been reported showing clear
signs of fluctuations in both, the real and imaginary parts of the
$\it{ac}$
conductivity.\cite{Anlage:96,Booth:96,Nakielski:97,Neri:98,Waldram:99}
The real part $\sigma_1$ of the complex conductivity
$(\displaystyle \widetilde{\sigma} = \sigma_1 - i \sigma_2)$ has a
sharp peak at $T_c$, which is not observed in e.g. Nb as a
representative of low temperature classical superconductors.
\cite{Klein:94} The salient feature of the $\it{ac}$ case is that
the fluctuation conductivity does not diverge at $T_c$ because a
finite frequency provides a limit to the observation of the
critical slowing down near $T_c$. The real part $\sigma_1$ has a
maximum at $T_c$. It is also important to note that $\sigma_1$ and
$\sigma_2$ have individually different temperature and frequency
dependences, even though they result from the same underlying
physics. Testing a given theoretical model becomes more stringent
when two curves have to be fitted with the same set of parameters.
\par
The expressions for the $\it{ac}$ fluctuation conductivity in the
Gaussian regime have been deduced within the time dependent
Ginzburg-Landau (TDGL) theory by Schmidt.\cite{Schmidt:68}
Using general physical arguments, Fisher, Fisher, and Huse
\cite{Fisher:91} provided a formulation for the scaling
of the complex $\it{ac}$ conductivity as
\begin{equation}
\widetilde{\sigma}(\omega) \propto
\displaystyle\xi^{z+2-D}\,\widetilde{{\cal S}}_{\pm}(\omega\,\xi^z) \,\,\, ,
\label{1-1}
\end{equation}
where $\xi$ is the correlation length, $z$ is the dynamical
critical exponent, $D$ is the dimensionality of the system, and
$\displaystyle \widetilde{{\cal S}}_{\pm}(\omega\xi^z)$ are some
complex scaling functions above and below $T_c$. This form of the
fluctuation conductivity was claimed to hold in both, the Gaussian
and critical regimes. Dorsey \cite{Dorsey:91} has deduced the
scaling functions in the Gaussian regime above $T_c$, and verified
the previous results of Schmidt.\cite{Schmidt:68} More recently,
Wickham and Dorsey \cite{Wickham:00} have shown that even in the
critical regime, where the quartic term in the GL free energy
plays a role, the scaling functions preserve the same form as in
the Gaussian regime.
\par
The above mentioned theoretical expressions of the $\it{ac}$ fluctuation
conductivity did not take into account the slow variation
approximation which is required for the validity of the
Ginzburg-Landau theory.\cite{Skocpol:75} It was noted long time
ago that the summation over the fluctuation modes had to be
truncated at a wavevector which corresponded roughly to the
inverse of the intrinsic coherence length $\xi_0$.\cite{Patton:69}
The improved treatment with a short wavelegth cutoff was
applied in fluctuation diamagnetism,\cite{Gollub:73} and
$\it{dc}$ paraconductivity far above $T_c$.\cite{Johnson:78}
This approach was also applied in $\it{dc}$ fluctuation conductivity of
high-$T_c$ superconductors where one encounters a large
anisotropy.\cite{Hopfengartner:91,Gauzzi:95,Silva:01}
The introduction of the short wavelength cutoff was found to be
essential in fitting the theoretical expressions to the
experimental data. In view of the great potential of the
microwave method described above, we find motivation to
elaborate in this paper the improved theory of $\it{ac}$ fluctuation
conductivity including the short wavelength cutoff.
We find that the resulting
expressions can be written in the form of Eq.~(\ref{1-1}).
However, the cutoff introduces a breakdown of the scaling
property in the variable $\omega \xi^z$. Also, we find that
the phase $\phi$ of the complex conductivity
($\tan{\phi} = \sigma_2/\sigma_1$) evaluated at $T_c$
departs from the value $\pi/4$ when cutoff is introduced.
Values of $\phi$ larger than $\pi/4$ were observed
experimentally,\cite{Booth:96} but were attributed to
an unusually large dynamic critical exponent.
Also, deviation of the scaling in the variable $\omega \xi^z$
was observed already at 2 K above $T_c$,\cite{Booth:96} but no
analysis was made considering the short wavelength cutoff
in the fluctuation spectrum. The present theory is developed
for different dimensionalities which facilitates comparison with
experimental data.

\section{THE EFFECTS OF SHORT WAVELENGTH CUTOFF}
\label{Sect2}

Frequency dependent conductivity can be calculated within the Kubo
formalism from the current correlation function. For the fluctuation
conductivity one has to consider the current due to the fluctuations
of the order parameter. The resulting expression for the real part of
the conductivity is \cite{Tinkham:95}
\begin{equation}
\sigma_1^{xx} = \left(\frac{2 e \hbar}{m}\right)^2 \frac{1}{k_B T}
\sum_{\bf k} k_x^2 <|\psi_{\bf k}|^2>^2 \frac{\tau_k /2}{1 +
(\omega\tau_k/2)^2} \,\,\, ,
\label{3-1}
\end{equation}
where the current is assumed to be in the $x$-direction. $\psi_k$
is the Fourier component of the order parameter, and
$\displaystyle \tau_k = \tau_0 /(1+\xi^2 k^2)$ is the relaxation
time of the $k$-th component. The relaxation time for the $k = 0$
mode is given by
\begin{equation}
\displaystyle
\tau_0 = \frac{\pi \hbar}{8 k_B T_c}\left(\frac{\xi(T)}
{\xi_0}\right)^z \,\,\, ,
\label{3-1a}
\end{equation}
where $z$ is the dynamic critical exponent. An alternative approach
is to calculate the response of the system to an external field
through the expectation value of the current operator averaged with respect to the
noise.\cite{Dorsey:91} However, the introduction of the short
wavelength cutoff in this approach leads only to selfconsistent
implicit expressions.\cite{Neri:99,Neri:99a}
\par
Eq.~(\ref{3-1}) is obtained from the time dependent
Ginzburg-Landau theory and represents the equivalent of the
Aslamazov-Larkin fluctuation conductivity obtained from
microscopic calculations. In the following, we present the results
which take account of the short wavelength cutoff in this
contribution to the ${\it ac}$ fluctuation conductivity. The other
contributions such as Maki-Thomson (MT) and one-electron density
of states (DOS)
renormalization\cite{Varlamov:80,Varlamov:97,Larkin:02} cannot be
treated within the time dependent Ginzburg-Landau theory but
require microscopic calculations. It has been
shown\cite{Varlamov:97} that MT anomalous contribution in
high-$T_c$ superconductors is almost temperature independent while
DOS contribution is strongly temperature dependent, and contains a
number of parameters which have to be determined through a complex
fitting procedure in an experimental data
analysis.\cite{Varlamov:99} Since the three terms in the
fluctuation conductivity are additive it is important to have the
Aslamazov-Larkin term corrected for short wavelength cutoff which
then allows to fit the MT and DOS contributions properly from the
rest of the total experimental fluctuation conductivity.
\par
The sum in Eq.~(\ref{3-1}) can be evaluated by integration
considering the appropriate dimensionality. In this section we
discuss the simplest case of an isotropic {\it 3D} superconductor.
The integration in  $\rm \bf k$-space needs a cutoff since the
order parameter cannot vary appreciably over distances which are
shorter than some minimum wavelength. The cutoff in $k_x$ can be
expressed as $\displaystyle k_x^{max} = \Lambda / \xi_0$, where
$\Lambda$ is a dimensionless cutoff parameter. Obviously,
$\displaystyle \Lambda \rightarrow \infty$ would imply no cutoff
in the integration, whereas for $\Lambda \approx 1$ one obtains
the usually assumed cutoff at $1/\xi_0$. In the {\it 3D} isotropic
case, the same cutoff applies also to $k_y$ and $k_z$ so that for
the {\it 3D} integration in the $\rm \bf k$-space one has to set
the cutoff limit for the modulus $\displaystyle k^{max} =
\sqrt{3}\Lambda / \xi_0$. With the change of variable $q(T) = k
\xi(T)$ one obtains
\begin{equation}
\displaystyle {\sigma_1}^{\hspace{-0.1cm}{\it
3D},iso}(\omega,T,\Lambda)=
\frac{e^2}{6\pi\hbar\xi_0}\left(\frac{\xi(T)}{\xi_0}\right)^{z-1}
\int_0^{Q} \frac{q^4}{(1+q^2)[\Omega^2 + (1+q^2)^2]}\,dq \,\,\, ,
\label{3-2}
\end{equation}
where
\begin{equation}
Q(T,\Lambda)=k_{cut}\,\xi(T)=\sqrt{3}\,\Lambda\left(\frac{\xi(T)}{\xi_0}\right)
\label{3-2a}
\end{equation}
is the temperature dependent cutoff limit in the $q$-space, and
\begin{equation}
\displaystyle \Omega(\omega,T)= \frac{\omega \tau_0}{2} =
\frac{\pi}{16}\frac{\hbar \omega}{k_B T_c}
\left(\frac{\xi(T)}{\xi_0}\right)^z
\label{3-3}
\end{equation}
is a dimensionless variable which depends on frequency and
temperature as independent experimental variables.
\par
For the $\it{dc}$ case ($\omega=0$), and no cutoff
($\displaystyle \Lambda \rightarrow \infty$), one finds from Eq.~(\ref{3-2})
\begin{equation}
\displaystyle {\sigma_{dc}}^{\hspace{-0.25cm}{\it
3D},iso}(T,\Lambda \rightarrow \infty)=\frac{e^2}{32\hbar\xi_0}
\left(\frac{\xi(T)}{\xi_0}\right)^{z-1} \,\,\, , \label{3-4}
\end{equation}
which reduces to the well known Aslamazov-Larkin result
\cite{Tinkham:95} provided that relaxational dynamics is assumed
($z = 2$), and $\displaystyle \xi(T)/ \xi_0$ is taken only in the
Gaussian limit as $1/\sqrt{\epsilon}$. However, with a finite
cutoff parameter $\Lambda$ one obtains
\begin{equation}
\displaystyle {\sigma_{dc}}^{ \hspace{-0.25cm}{\it
3D},iso}(T,\Lambda)= \frac{e^2}{16\pi\hbar\xi_0}
\left(\frac{\xi(T)}{\xi_0}\right)^{z-1}
\left[\arctan{(Q)}-Q\,\frac{\left(\displaystyle\frac{5}{3}\,Q^2+1\right)}
{(1+ Q^2)^2} \right] \,\,\, . \label{3-5}
\end{equation}
This result has been obtained by Hopfeng\"artner et. al.
\cite{Hopfengartner:91} except that they used only the Gaussian
limit $1/\sqrt{\epsilon}$ for the reduced correlation length
$\displaystyle \xi(T)/ \xi_0$. Their analysis has shown
that the cutoff plays no role exactly at $T_c$ since $Q
\rightarrow \infty$ regardless of $\Lambda$. However, at any
temperature above $T_c$ one gets a finite $Q$ and the value of the
conductivity is lowered with respect to the result given by
Eq.~(\ref{3-4}). Their conclusion was that the Gaussian fluctuations
with no cutoff yield an overestimated fluctuation conductivity.
\par
In this paper we are primarily interested in the ac case. Before
integrating Eq.~(\ref{3-2}) with $\Omega \not = 0$, we find the
corresponding expression for the imaginary part $\sigma_2$. We can
apply Kramers-Kronig relations to each of the Fourier components
in Eq.~(\ref{3-1}), and carry out the summation. This is
equivalent to a calculation of the kernel ${\cal K}_2$ for
$\sigma_2$ from the kernel ${\cal K}_1$ used in Eq.~(\ref{3-2}),
namely
\begin{equation}
{\cal K}_2(\Omega) = \frac{2\Omega}{\pi}\int_0^{\infty}
\frac{{\cal K}_1(\Omega')}{\Omega^2-{\Omega'}^2}\,
d\Omega' \,\,\, .
\label{3-6}
\end{equation}
With the kernel $\displaystyle {\cal K}_2 (\Omega)$, the imaginary
part of the fluctuation conductivity can be calculated for any cutoff
parameter $\Lambda$
\begin{equation}
\displaystyle{\sigma_2}^{ \hspace{-0.1cm}{\it
3D},iso}(\omega,T,\Lambda)= \frac{e^2}{6\pi\hbar\xi_0}
\left(\frac{\xi(T)}{\xi_0}\right)^{z-1}
\int_0^{Q}\frac{\Omega\,q^4}{(1+q^2)^2[\Omega^2 + (1+ q^2)^2]}\,dq
\,\,\, . \label{3-7}
\end{equation}
Finally, the complex fluctuation conductivity can be written in
the form
\begin{equation}
\displaystyle {\widetilde\sigma}^{{\it 3D},iso}(\omega,T,\Lambda)=
\frac{e^2}{32 \hbar \xi_0} \left(\frac{\xi(T)}{\xi_0}\right)^{z-1}
\left[{{\cal S}_1}^{\hspace{-0.1cm}{\it 3D},iso}(\omega,T,\Lambda)
+ i\,{{\cal S}_2}^{\hspace{-0.1cm}{\it
3D},iso}(\omega,T,\Lambda)\right] \,\,\, . \label{3-8}
\end{equation}
The prefactor is equal to the $\it{dc}$ result with no cutoff effect as
in Eq.~(\ref{3-4}). The functions $\displaystyle {\cal S}_{1,2}$
are given by the following  expressions
\begin{equation}
\displaystyle {\cal S}_1^{\hspace{0.05cm}{\it
3D},iso}(\omega,T,\Lambda) =
\frac{1}{3\pi\Omega^2}\left[P_{-}(P_{+}^2+2)L+2\,P_{+}(P_{-}^2-2)A+
16\,\arctan{(Q)}\right] \,\,\, , \label{3-9}
\end{equation}
\begin{equation}
\displaystyle {\cal S}_2^{\hspace{0.05cm}{\it
3D},iso}(\omega,T,\Lambda) =
\frac{1}{3\pi\Omega^2}\left[2\,P_{-}(P_{+}^2+2)A-P_{+}(P_{-}^2-2)L-
24\,\Omega\,\arctan{(Q)}+8\,\Omega\,\frac{Q}{1+Q^2}\right] \,\,\,
, \label{3-10}
\end{equation}
where we used the following shorthand notations
\begin{equation}
\displaystyle P_{\pm} = \sqrt{2}\sqrt{\sqrt{\Omega^2+1}\pm 1} \,\,\, ,
\label{3-11}
\end{equation}
\begin{equation}
\displaystyle L =
\ln{\left(\frac{2+Q^2+(Q-P_{-})^2}{2+Q^2+(Q+P_{-})^2}\right)} \,\,\, ,
\label{3-12}
\end{equation}
\begin{equation}
\displaystyle A = \arctan{\left(\frac{2\,Q+P_{-}}{P_{+}}\right)}+
\arctan{\left(\frac{2\,Q - P_{-}}{P_{+}}\right)} \,\,\, .
\label{3-13}
\end{equation}
It can be easily verified that the
$\displaystyle {\cal S}_{1,2}$-functions given by Eq.~(\ref{3-9}) and
Eq.~(\ref{3-10}) have proper limits. In the $\it{dc}$ limit
($\displaystyle \Omega \rightarrow 0$), one finds that
$\displaystyle {\cal S}_2 \rightarrow 0$, and $\displaystyle {\cal S}_1$
leads to the $\it{dc}$ result of Eq.~(\ref{3-5}). One can also verify that
the $\it{ac}$ results obtained previously by Schmidt
\cite{Schmidt:68} and Dorsey \cite{Dorsey:91}
can be recovered from our Eq.~(\ref{3-9}) and Eq.~(\ref{3-10}) in
the limit $\displaystyle \Lambda \rightarrow \infty$, i.~e. when
no cutoff is made.
\par
\begin{figure}
\centerline{\includegraphics[width=0.8\textwidth]{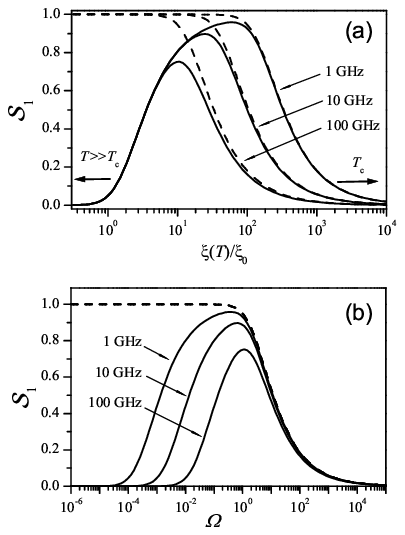}}
\caption{${\cal S}_1$-curves for the {\it 3D} isotropic case
calculated from Eq.~(\ref{3-9}) for a finite cutoff parameter
$\Lambda = 0.5$ (full lines), and for $\Lambda = \infty$ (dashed
lines). The variable is $\displaystyle \xi(T)/ \xi_0$ in (a), and
$\Omega$ in (b). The curves are labelled by the frequency
$\displaystyle \omega/2 \pi$ used in the calculations.}
\label{Fig1}
\end{figure}

The effects of the cutoff are not trivial in the $\it{ac}$ case.
It is essential to examine those effects in detail as they have
strong bearing on the analysis of the experimental data. The
prefactor in Eq.~(\ref{3-8}) depends only on temperature while the
cutoff parameter $\Lambda$ is found only in the $\cal
S$-functions. Therefore, the effects of the cutoff can be studied
through the $\cal S$-functions alone. We can look at the
temperature and frequency dependences of these functions with and
without the cutoff. Fig.~\ref{Fig1}(a) shows a set of ${\cal S}_1$
curves as functions of $\displaystyle \xi(T)/ \xi_0$ for three
different frequencies. Far above $T_c$ the relaxation time
$\tau_0$ is so short that $\omega \tau_0 \ll 1$ for any of the
chosen frequencies. Therefore the response of the system is like
in the $\it{dc}$ case. With no cutoff, ${\cal S}_1$ saturates to
unity (dashed lines in Fig.~\ref{Fig1}(a)). This limit is required
in order that $\sigma_1$ from Eq.~(\ref{3-8}) becomes equal to
$\displaystyle \sigma_{dc} (\Lambda \rightarrow \infty)$ in
Eq.~(\ref{3-4}). If a cutoff with a finite $\Lambda$ is included,
${\cal S}_1$ decays at higher temperatures (solid lines in
Fig.~\ref{Fig1}(a)). The reduction of ${\cal S}_1$ is more
pronounced at smaller values of $\displaystyle \xi(T)$ since the
integration in the $q$-space is terminated at a lower value
$\displaystyle Q = \sqrt{3}\ \Lambda\ \xi(T)/\xi_0$. At higher
temperatures the conductivity $\sigma_1$ at any frequency behaves
asymptotically as $\sigma_{dc}$ given by Eq.~(\ref{3-5}).
\par
At temperatures closer to $T_c$ the relaxation time $\tau_0$
increases as $\xi(T)^z$ with increasing correlation length
according to Eq.~(\ref{3-1a}) which is usually termed
critical slowing down. When
$\displaystyle \omega \tau_0 \approx 1$ for a given frequency,
${\cal S}_1$ is sharply reduced and vanishes in the limit of $T_c$.
With the diverging prefactor in Eq.~(\ref{3-8}) it can still yield a
finite $\sigma_1$ at $T_c$. It may appear from Fig.~\ref{Fig1}(a) that
cutoff makes no effect when $T_c$ is approached, but we show later
that an important feature still persists in $\sigma_1$.
\par
Obviously, at lower operating frequencies one needs to approach
$T_c$ closer so that the critical slowing down could reach the
condition $\displaystyle \omega \tau_0 \approx 1$. One can see
from Fig.~\ref{Fig1}(a) that for frequencies below 1 GHz one would
have to approach $T_c$ closer than 1 mK in order to probe the
critical slowing down in fluctuations. The higher the frequency,
the farther above $T_c$ is the temperature where the crossover
$\displaystyle \omega \tau_0 \approx 1$ occurs. This feature
expresses the scaling property of the conductivity in frequency
and temperature variables. However, the scaling property
holds strictly only in the absence of the cutoff. Namely, if one
sets $\Lambda \rightarrow \infty$, the function ${\cal S}_1$ depends
only on the scaling variable $\Omega$. Fig.~\ref{Fig1}(b) shows
the same set of curves as in Fig.~\ref{Fig1}(a), but plotted
versus $\Omega$. The three dashed curves from Fig.~\ref{Fig1}(a)
coalesce into one dashed curve in Fig.~\ref{Fig1}(b), thus showing
the scaling property in the absence of cutoff. However, the full
lines representing the functions ${\cal S}_1$ with a finite cutoff
parameter $\Lambda$ do not scale with the variable $\Omega$. The
reason is that the function ${\cal S}_1$ then depends also on $Q$,
which itself is not a function of $\Omega$. Namely, the cutoff in
the $q$-space depends on the properties of the sample, and on the
temperature, but not on the frequency used in the experiment.
Hence, the cutoff brings about a breakdown of the scaling property
in frequency and temperature. The effect is more pronounced at
temperatures farther above $T_c$ where the cutoff is stronger.
\par
\begin{figure}
\centerline{\includegraphics[width=0.8\textwidth]{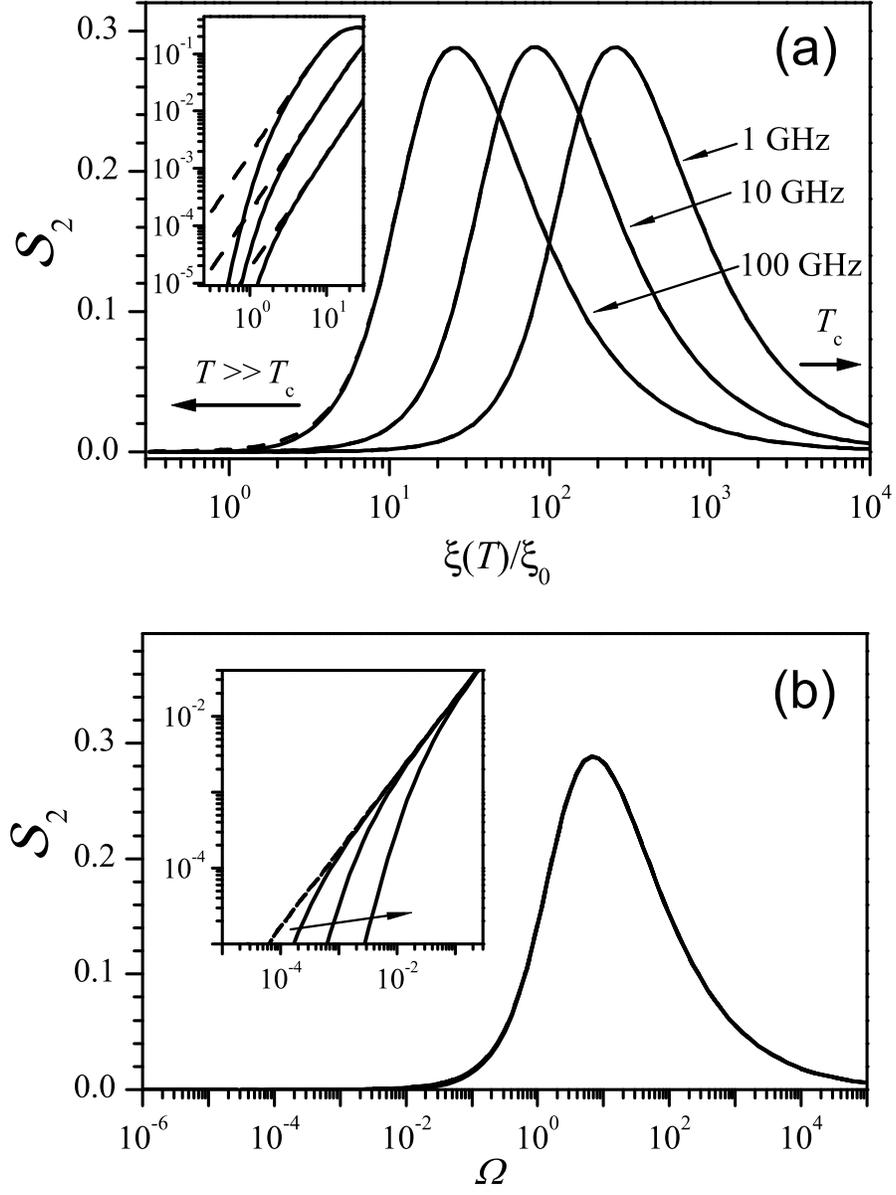}}
\caption{${\cal S}_2$-curves for the {\it 3D} isotropic case
calculated from Eq.~(\ref{3-10}) for the same parameters as in
Fig.~\ref{Fig1}. The effects of a finite cutoff are small and can
be seen only on logarithmic scales used in the insets.}
\label{Fig2}
\end{figure}

The properties of the function ${\cal S}_2$ are shown in
Fig.~\ref{Fig2} for the same set of three measurement frequencies
as in Fig.~\ref{Fig1}. When plotted versus $\displaystyle \xi(T)/ \xi_0$,
the function ${\cal S}_2$ exhibits a
maximum at the point where the corresponding function ${\cal S}_1$
shows the characteristic crossover due to $\displaystyle \omega
\tau_0 \approx 1$ as discussed above. When $T_c$ is approached,
${\cal S}_2$ tends to zero. When ${\cal S}_2$ is multiplied with the
diverging prefactor in Eq.~(\ref{3-8}), one finds a finite
$\sigma_2$ at $T_c$. Far above $T_c$, the function ${\cal S}_2$
vanishes, regardless of the cutoff. This is consistent with the
behavior of ${\cal S}_1$. Namely, at high enough
temperatures, ${\cal S}_1$ acquires asymptotically the $\it{dc}$ value,
as seen in Fig.~\ref{Fig1}.
Obviously, the imaginary part of the conductivity must
vanish when $\it{dc}$ like limit is approached.
The decrease of the function ${\cal S}_2$ at higher
temperatures is very rapid so that the effects of the cutoff are
unnoticeable on the linear scale. Only with the logarithmic scale
used in the inset to Fig.~\ref{Fig2}(a), one observes that the
cutoff effects are present also in ${\cal S}_2$, though by a very
small amount. Fig.~\ref{Fig2}(b) shows the scaling property of
${\cal S}_2$ with no cutoff and its breakdown when cutoff is
included.
\par
\begin{figure}
\centerline{\includegraphics[width=0.8\textwidth]{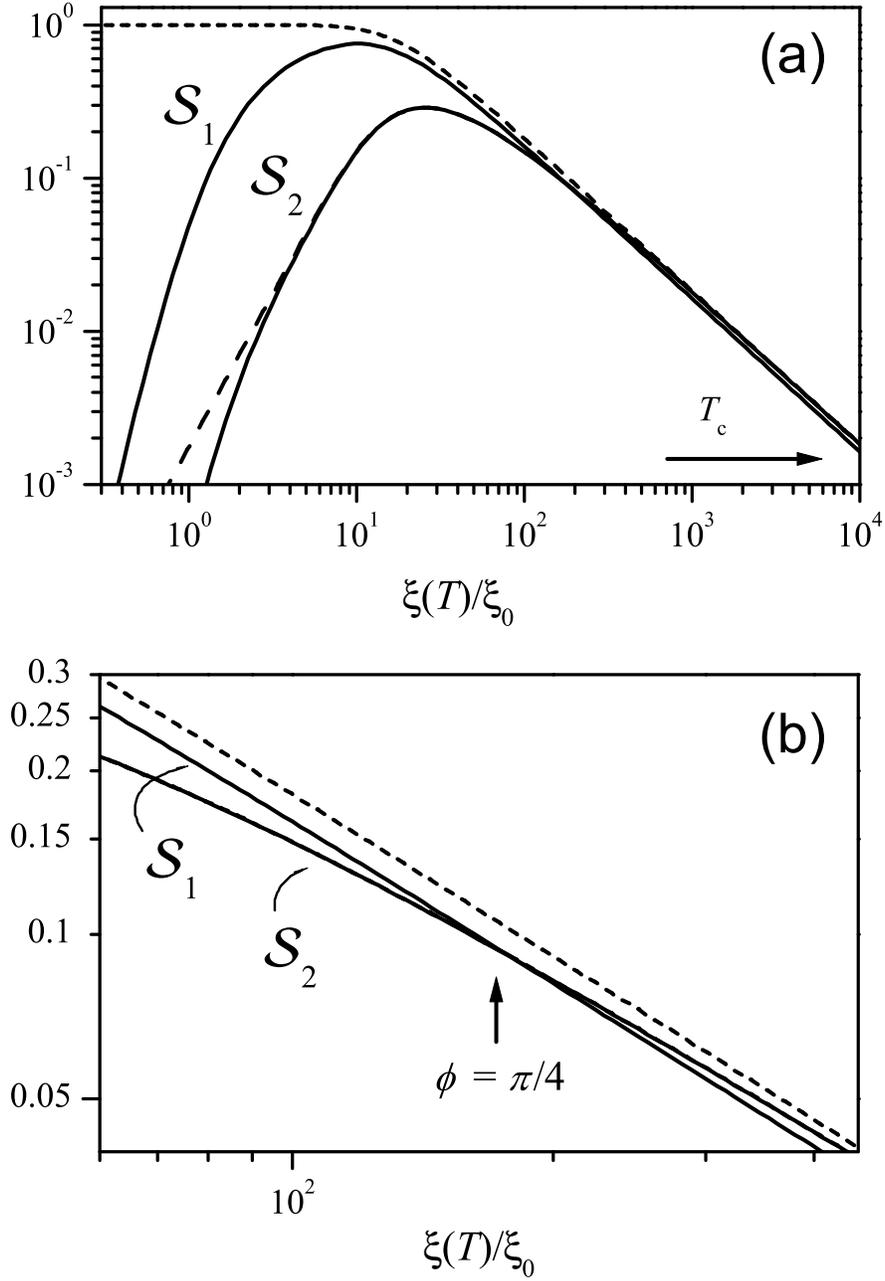}}
\caption{(a) Asymptotic behavior of the functions ${\cal S}_1$ and
${\cal S}_2$ of the {\it 3D} isotropic case calculated with the
choice $\displaystyle \omega/2 \pi = 100$ GHz in the limit $T
\rightarrow T_c$. The dashed lines are the ${\cal S}_{1,2}$-curves
calculated with no cutoff ($\Lambda = \infty$), and the full lines
include a finite cutoff ($\Lambda = 0.5$). (b) Enlarged section
which shows the crossing of ${\cal S}_1$ and ${\cal S}_2$ ($\phi =
\pi/4$, see text) at a temperature slightly above $T_c$. The two
dashed lines are indistinguishable in this temperature range.}
\label{Fig3}
\end{figure}

We have noted above that both, ${\cal S}_1$ and ${\cal S}_2$ tend
to zero when $T_c$ is approached. Also, the effect of cutoff is
seen to be small in that limit. Yet, these functions are
multiplied by the diverging prefactor in Eq.~(\ref{3-8}), and then
may yield finite $\sigma_1$ and $\sigma_2$. A careful analysis is
needed in order to find the phase $\phi$ of the complex
conductivity ($\displaystyle \phi = \arctan{(\sigma_2 /
\sigma_1)}$) at $T_c$. For a {\it 3D} isotropic superconductor,
Dorsey \cite{Dorsey:91} has predicted $\displaystyle \phi = \pi/
4$, i.~e. $\displaystyle \sigma_1 = \sigma_2$ at $T_c$. His result
was obtained with no cutoff and it remains to be seen if this
property is preserved even when a finite cutoff is made.
Fig.~\ref{Fig3}(a) shows ${\cal S}_1$ and ${\cal S}_2$ as
functions of $\displaystyle \xi(T)/\xi_0$ for 100 GHz frequency.
The effects of cutoff on ${\cal S}_2$ are noticeable only far
above $T_c$. Closer to $T_c$, the curves for ${\cal S}_2$
calculated with, and without cutoff, are indistinguishable. In
contrast, a finite cutoff reduces the values of ${\cal S}_1$ even
in the limit of $T_c$. As a result, the final cutoff parameter
$\Lambda$ yields a crossing of the curves for ${\cal S}_1$ and
${\cal S}_2$ at some temperature slightly above $T_c$. It is
better seen on an enlarged scale in Fig.~\ref{Fig3}(b). This is a
surprising result which has bearing on the experimental
observations. Due to the cutoff, the condition $\displaystyle
{\cal S}_1={\cal S}_2$ ($\displaystyle \phi = \pi/4$) is reached
at a temperature slightly above $T_c$. Since both, ${\cal S}_1$
and ${\cal S}_2$ are multiplied with the same prefactor in
Eq.~(\ref{3-8}), one finds that the crossing of $\displaystyle
\sigma_1 (T)$ and $\displaystyle \sigma_2 (T)$ does not occur at
the peak of $\displaystyle \sigma_1 (T)$, but at a slightly higher
temperature. Exactly at $T_c$, $\sigma_2$ is higher than
$\sigma_1$ since a finite cutoff parameter reduces $\sigma_1$, but
makes no effect on $\sigma_2$.
\par
\begin{figure}
\centerline{\includegraphics[width=0.8\textwidth]{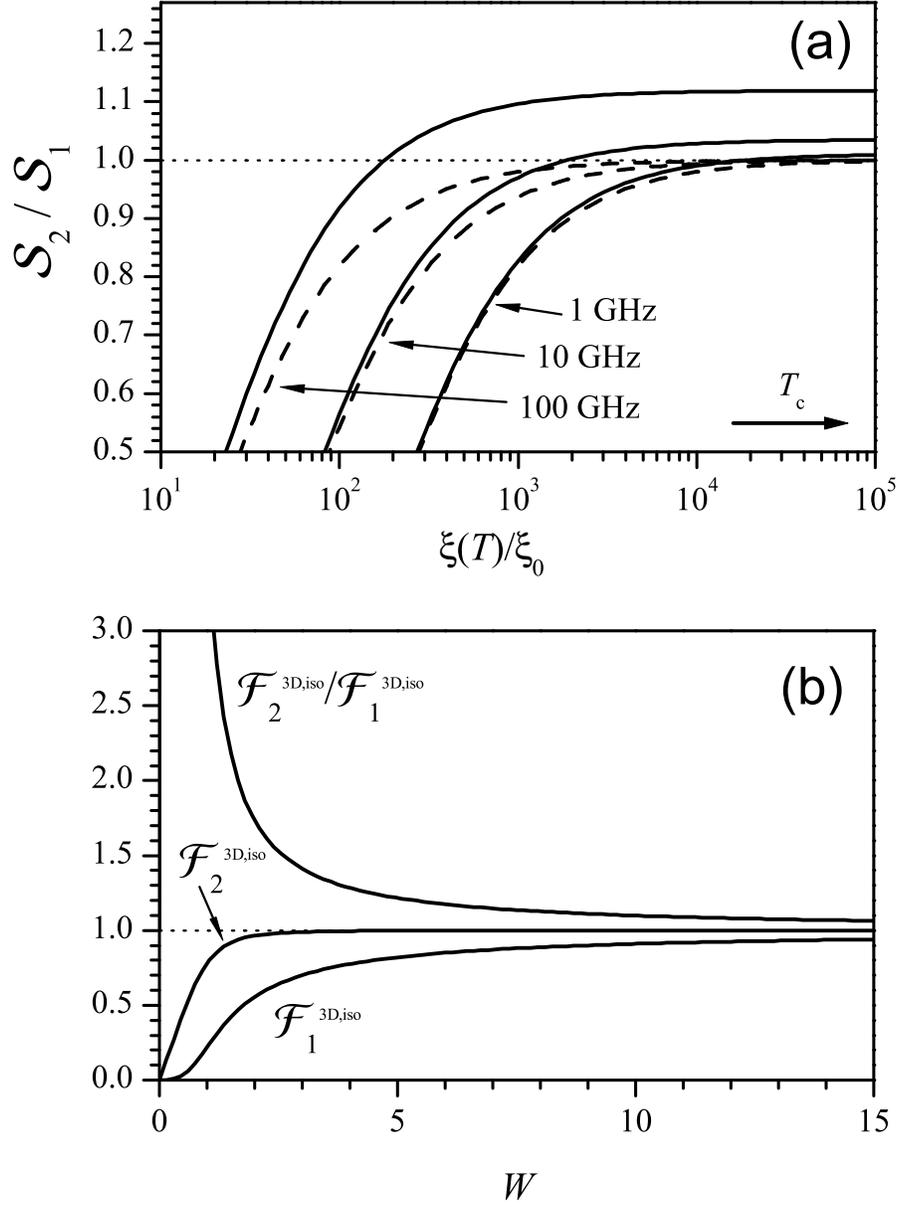}}
\caption{(a)
The ratio $\displaystyle {\cal S}_2/{\cal S}_1$ (equal to
$\displaystyle\sigma_2/\sigma_1$) for the {\it 3D} isotropic case
at temperatures approaching $T_c$. With no cutoff (dashed lines)
the ratio tends to unity for all frequencies. With a finite cutoff
($\Lambda = 0.5$) the ratio equals unity at a temperature slightly
above $T_c$, dependent on the frequency. In the limit $T
\rightarrow T_c$ the ratio saturates to a frequency dependent
value larger than unity. (b) The ratio $\displaystyle {\cal
S}_2/{\cal S}_1$ and ${\cal F}_{1,2}$ given by Eq.~(\ref{3-20}).
The variable $W$ is defined by Eq.~(\ref{3-18}).} \label{Fig4}
\end{figure}

The observation that the cutoff brings about a reduction of
$\sigma_1$ at $T_c$ is worth further investigation since it can be
measured experimentally. Fig.~\ref{Fig4}(a) shows the ratio
$\displaystyle {\cal S}_2 / {\cal S}_1$ (equal to $\displaystyle
\sigma_2 / \sigma_1$) at temperatures approaching $T_c$. With no
cutoff (dashed lines in Fig.~\ref{Fig4}(a)), this ratio reaches
unity regardless of the frequency used. A finite cutoff parameter
($\displaystyle \Lambda = 0.5$ in Fig.~\ref{Fig4}(a)) makes the
ratio equal to unity at a temperature slightly above $T_c$, and in
the limit of $T_c$ the ratio saturates at some higher value. The
saturation level is seen to be higher when a higher frequency is
used.
\par
One can find analytical expansions of the $\cal S$-functions in
the limit of $T_c$ ($\Omega \rightarrow \infty$). The leading
terms are
\begin{equation}
\displaystyle {\cal S}_{1,2}^{\hspace{0.02cm}{\it
3D},iso}(W,\Omega \rightarrow \infty) \approx \frac{4 \sqrt{2}}{3
\pi}\left(C\mp \frac{1}{2}\, D\right)\frac{1}{\sqrt{\Omega}}
\,\,\, , \label{3-14}
\end{equation}
where we used the notation
\begin{equation}
\displaystyle C = \arctan{(1+\sqrt{2}W)}-\arctan{(1-\sqrt{2}W)} \,\,\, ,
\label{3-16}
\end{equation}
\begin{equation}
\displaystyle
D = \ln{\left(\frac{1+\sqrt{2}W+W^2}{1-\sqrt{2}W+W^2}\right)} \,\,\, .
\label{3-17}
\end{equation}
The parameter $W$ depends on the frequency $\omega$ and the cutoff
parameter $\Lambda$
\begin{equation}
\displaystyle W = \sqrt{3}\,\Lambda\,
\sqrt{\frac{16}{\pi}\frac{k_B T_c}{\hbar\omega}} \,\,\, .
\label{3-18}
\end{equation}
Both functions tend to zero in the limit of $T_c$ ($\displaystyle
\Omega \rightarrow \infty$), but their ratio is finite and depends
on the parameter $W$. Fig.~\ref{Fig4}(b) shows the plot of
$\displaystyle {\cal S}_2 /{\cal S}_1$ at $T_c$ as the function of $W$. One can
observe that for a given cutoff parameter $\Lambda$, the ratio
$\displaystyle {\cal S}_2 /{\cal S}_1$ at $T_c$ increases at
higher frequencies (lower $W$). The limits at $T_c$ in
Fig.~\ref{Fig4}(a) represent only three selected points on the
curve for $\displaystyle {\cal S}_2 /{\cal S}_1$ in
Fig.~\ref{Fig4}(b).
\par
In a given experiment, the ratio $\displaystyle \sigma_2 /
\sigma_1$ at $T_c$ can be directly determined from the
experimental data so that the corresponding value of the parameter
$W$ can be found uniquely from the curve of $\displaystyle {\cal
S}_2 /{\cal S}_1$ in Fig.~\ref{Fig4}(b), and the cutoff parameter
$\Lambda$ is obtained using Eq.~(\ref{3-18}). We should note that
$\Lambda$ is a temperature independent parameter. It can be
determined by the above procedure from the experimental data at
$T_c$, but it controls the cutoff at all temperatures.
\par
One may observe from Eq.~(\ref{3-14}) that in the limit of $T_c$
the leading terms in the expansions of the $\cal S$-functions
behave as $(\displaystyle \xi(T)/\xi_0)^{-z/2}$. Taking into
account the prefactor in Eq.~(\ref{3-8}) one finds that $\sigma_1$
and $\sigma_2$ can have finite nonzero values at $T_c$ only if $z
= 2$, i.~e. for the purely relaxational dynamics. We have assumed
this case in all the figures of this section.
\par
From the experimental data at $T_c$ one can determine also the
parameter $\xi_0$. Using Eq.~(\ref{3-8}) and the ${\cal S}$-functions
in Eq.~(\ref{3-14}), one obtains finite conductivities at $T_c$
\begin{equation}
\displaystyle {\sigma}_{1,2}^{\hspace{0.02cm}{\it
3D},iso}(\omega,T_c,\Lambda)=
\frac{e^2}{6\hbar\xi_0}\sqrt{\frac{2}{\pi}\frac{k_B
T_c}{\hbar\omega}} \, {\cal F}_{1,2}^{\hspace{0.02cm}{\it
3D},iso}(W) \,\,\, , \label{3-19}
\end{equation}
where
\begin{equation}
\displaystyle {\cal F}_{1,2}^{\hspace{0.02cm}{\it 3D},iso}(W)=
\frac{1}{\pi}\left(C\,\mp\,\frac{1}{2}\,D\right) \,\,\, .
\label{3-20}
\end{equation}
As explained above, from the ratio of the experimental values
$\displaystyle \sigma_2 / \sigma_1$ at $T_c$ one can determine the
parameter $W$, and the values of $\displaystyle {\cal F}_{1,2}$
can then be calculated from Eq.~(\ref{3-20}). The remaining unknown
parameter $\xi_0$ can be obtained using Eq.~(\ref{3-19}) and
either of the experimental values of $\displaystyle \sigma_1 (T_c)$
or $\displaystyle\sigma_2 (T_c)$.
\par
It is also interesting to look at the plots of $\displaystyle
{\cal F}_{1,2} (W)$ in Fig.~\ref{Fig4}(b). One can observe that
${\cal F}_2$ saturates to unity already at small values of
$\displaystyle W \ge 2$. On the other hand, ${\cal F}_1$ is
smaller than unity at any finite value of $W$, in conformity with
the ratio $\displaystyle ({\cal F}_2 / {\cal F}_1 )=({\cal S}_2
/{\cal S}_1 )$ at $T_c$. At this point it is useful to find the
expected range of the values of $W$ encountered in the
experiments. For the microwave frequencies in the range 1-100 GHz,
with $\displaystyle \Lambda = 0.5$, and $\displaystyle T_c = 100
K$, one finds that $W$ is in the range 9 - 90. According to
Fig.~\ref{Fig4}(b), $\displaystyle {\cal F}_2 \approx 1$ in this
range. This means that the cutoff makes no effect on ${\cal S}_2$,
and only ${\cal S}_1$ is reduced, in conformity with the
calculated curves shown in Fig.~\ref{Fig3}.
\section{ANISOTROPY}
\label{Sect3}
Most high-$T_c$ superconductors are anisotropic, some of them even
having a high value of the anisotropy parameter $\displaystyle
\gamma = \xi_{ab0} /\xi_{c0}$. Therefore, for practical purposes
one needs adequate expressions for the $\it{ac}$ fluctuation
conductivity. The real part of the fluctuation conductivity in the
ab-plane is obtained using the Kubo formalism as in the isotropic
case. One obtains
\begin{equation}
\displaystyle {\sigma_1}^{\hspace{-0.1cm}{\it 3D},aniso}=
\left(\frac{2\,e\,\hbar}{m_{ab}\,\alpha}\right)^2\, k_B\,T_c\,
\sum_{\bf k}
\frac{k_x^2\,\tau_0/2}{(1+k_{ab}^2\,\xi_{ab}^2+k_c^2\,\xi_c^2)
\left[\Omega^2+(1+k_{ab}^2\,\xi_{ab}^2+k_c^2\,\xi_c^2)^2\right]}
\,\,\, . \label{4-1}
\end{equation}
Taking $\displaystyle k_x^2 = k_{ab}^2/2$, and substituting the
variables $\displaystyle q_{ab} = k_{ab}\xi_{ab}$ and
$\displaystyle q_c = k_c\xi_c$, one can evaluate the sum
in Eq.~(\ref{4-1}) by integration in the $q_{ab}$-plane,
and along the $q_c$-axis
\begin{equation}
\displaystyle {\sigma_1}^{\hspace{-0.1cm}{\it 3D},aniso}
(\omega,T,\Lambda_{ab},\Lambda_c)=
\frac{e^2}{8\pi\hbar\xi_{0c}}\left(\frac{\xi(T)}{\xi_0}\right)^{z-1}
\int_0^{Q_{ab}}\int_{-Q_c}^{Q_c}
\frac{q_{ab}^{\,3}}{(1+q_{ab}^2+q_c^2)[\Omega^2+(1+q_{ab}^2+q_c^2)^2]}
\,d q_{ab}\,d q_c \,\,\, , \label{4-2}
\end{equation}
where we allowed a cutoff $\displaystyle Q_{ab}(T) =
\sqrt{2} \Lambda_{ab}\,\xi_{ab}(T)/ \xi_{0ab}$ in the
$q_{ab}$-plane, and a possibly different cutoff
$\displaystyle Q_c(T) = \Lambda_c\,\xi_c(T)/ \xi_{0c}$ along
the $q_c$-axis. The dimensionless parameter $\Omega$ is the same
as given by Eq.~(\ref{3-3}). We use the notation
$\displaystyle \xi(T)/\xi_0$ for both
$\displaystyle \xi_{ab}(T)/ \xi_{0ab}$ and
$\displaystyle \xi_c(T)/ \xi_{0c}$.
\par
We may briefly examine the $\it{dc}$ case ($\displaystyle \Omega = 0$).
With no cutoff one obtains
\begin{equation}
\displaystyle {\sigma_{dc}}^{\hspace{-0.25cm}{\it 3D},aniso}
(T,\Lambda_{ab,c}\rightarrow \infty)=
\frac{e^2}{32\hbar\xi_{0c}}\left(\frac{\xi(T)}{\xi_0}\right)^{z-1}
\,\,\, , \label{4-3}
\end{equation}
which reduces to the Aslamazov-Larkin result for $z=2$ (relaxational
dynamics) and $\displaystyle \xi(T)/\xi_0$ taken in the Gaussian
limit. Note that the fluctuation conductivity in the $ab$-plane
depends on $\xi_{0c}$. Finite cutoff parameters reduce the fluctuation
conductivity when the temperature is increased above $T_c$
\begin{eqnarray}
\displaystyle {\sigma_{dc}}^{\hspace{-0.25cm}{\it
3D},aniso}(T,\Lambda_{ab},\Lambda_c)=
\frac{e^2}{16\pi\hbar\xi_{0c}}\left(\frac{\xi(T)}{\xi_0}\right)^{z-1}&
\displaystyle
\left[\arctan{(Q_c)}-\frac{Q_{ab}^2\,Q_c}{2\,(1+Q_{ab}^2)\,
(1+Q_{ab}^2+Q_{c}^2)}\right.
\label{4-4} \\
\nonumber &\left.\displaystyle -\frac{2+3\,Q_{ab}^2}{2\,(1+Q_{ab}^2)^{3/2}}\,
\arctan{\left(\frac{Q_c}{\sqrt{1+Q_{ab}^2}}\right)}\right] \,\,\, .
\end{eqnarray}
This expression has not been reported in the previous literature.
The analysis of a $\it{dc}$ fluctuation conductivity is difficult because of
the number of unknown parameters.
\par
The $\it{ac}$ fluctuation conductivity can be obtained from the integral in
Eq.~(\ref{4-2}) for the real part while the imaginary part is obtained
by the procedure analogous to that of the isotropic case described in
the preceding section
\begin{equation}
\displaystyle {\sigma_2}^{\hspace{-0.1cm}{\it 3D},aniso}
(\omega,T,\Lambda_{ab},\Lambda_c)=
\frac{e^2}{8\pi\hbar\xi_{0c}}\left(\frac{\xi(T)}{\xi_0}\right)^{z-1}
\int_0^{Q_{ab}}\int_{-Q_c}^{Q_c}\frac{\Omega\,q_{ab}^{\,3}}
{(1+q_{ab}^2+q_c^2)^2[\Omega^2+(1+q_{ab}^2+q_c^2)^2]}\, d
q_{ab}\,d q_c \,\,\, . \label{4-4a}
\end{equation}
The full expression can again be written in the form
\begin{equation}
\displaystyle {\widetilde\sigma}^{{\it
3D},aniso}(\omega,T,\Lambda_{ab},\Lambda_c)= \frac{e^2}{32 \hbar
\xi_{0c}}\left(\frac{\xi(T)}{\xi_0}\right)^{z-1} \left[{{\cal
S}_1}^{\hspace{-0.1cm}{\it
3D},aniso}(\omega,T,\Lambda_{ab},\Lambda_c) +i\,{{\cal
S}_2}^{\hspace{-0.1cm}{\it 3D},aniso}
(\omega,T,\Lambda_{ab},\Lambda_c)\right] \,\,\, . \label{4-5}
\end{equation}
The $\cal S$-functions for the {\it 3D} anisotropic case are found
to be
\begin{eqnarray}
\displaystyle \hspace{-1cm}{\cal S}_1^{\hspace{0.05cm}{\it
3D},aniso} (\omega,T,\Lambda_{ab},\Lambda_c)= \frac{1}{3 \pi
\Omega^2}& \left[2\,Q_c(3+Q_c^2)\,L_1+P_{-}(P_{+}^2+2)\,L_2-
T_{-}(T_{+}^2+2-Q_{ab}^2)\,L_3+2\,P_{+}(P_{-}^2-2)\,A_1\right.
\label{4-6} \\
\nonumber &\left. \displaystyle  -2\,T_{+}(T_{-}^2-2+Q_{ab}^2)\,A_2-
8\,\sqrt{1+Q_{ab}^2}\,(2-Q_{ab}^2)\,A_3+16\,A_4
+12\,\Omega\,Q_c\,A_5\right] \,\,\, ,
\end{eqnarray}
\begin{eqnarray}
\displaystyle \hspace{-1cm}{\cal S}_2^{\hspace{0.05cm}{\it
3D},aniso} (\omega,T,\Lambda_{ab},\Lambda_c)= \frac{1}{3 \pi
\Omega^2}& \left[4\,Q_c\,(3+Q_c^2)\,A_5+2\,P_{-}(P_{+}^2+2)\,A_1-
2\,T_{-}(T_{+}^2+2-Q_{ab}^2)\,A_2-P_{+}(P_{-}^2-2)\,L_2\right.
\label{4-7} \\
\nonumber &\left. \displaystyle +T_{+}(T_{-}^2-2+Q_{ab}^2)\,L_3+
12\,\Omega\,\frac{2+Q_{ab}^2}{\sqrt{1+Q_{ab}^2}}\,A_3-24\,\Omega\,A_4-
6\,\Omega\,Q_c\,L_1\right] \,\,\, ,
\end{eqnarray}
where we used the shorthand notations for $\displaystyle P_{\pm}$ as
in Eq.~(\ref{3-11}), and the following
\begin{equation}
\displaystyle T_{\pm} =
\sqrt{2}\,\sqrt{\sqrt{(1+Q_{ab}^2)^2+\Omega^2}\,\pm(1+Q_{ab}^2)} \,\,\, ,
\label{4-8}
\end{equation}
\begin{equation}
\displaystyle L_1 =
\ln{\left(\frac{(1+Q_c^2)^2\,[\Omega^2+(1+Q_{ab}^2+Q_c^2)^2]}
{(1+Q_{ab}^2+Q_c^2)^2\,[\Omega^2+(1+Q_c^2)^2]}\right)} \,\,\, ,
\label{4-9}
\end{equation}
\begin{equation}
\displaystyle L_2 =
\ln{\left(\frac{2+Q_c^2+(Q_c-P_{-})^2}{2+Q_c^2
+(Q_c+P_{-})^2)}\right)} \,\,\, ,
\label{4-10}
\end{equation}
\begin{equation}
\displaystyle L_3 =
\ln{\left(\frac{2\,(1+Q_{ab}^2) + Q_c^2 +(Q_c-T_{-})^2}
{2\,(1+Q_{ab}^2) + Q_c^2 + (Q_c+T_{-})^2}\right)} \,\,\, ,
\label{4-11}
\end{equation}
\begin{equation}
\displaystyle A_1 = \arctan{\left(\frac{2\,Q_c + P_{-}}{P_{+}}\right)}+
\arctan{\left(\frac{2\,Q_c-P_{-}}{P_{+}}\right)} \,\,\, ,
\label{4-12}
\end{equation}
\begin{equation}
\displaystyle A_2 = \arctan{\left(\frac{2\,Q_c + T_{-}}{T_{+}}\right)}+
\arctan{\left(\frac{2\,Q_c-T_{-}}{T_{+}}\right)} \,\,\, ,
\label{4-13}
\end{equation}
\begin{equation}
\displaystyle A_3 =
\arctan{\left(\frac{Q_c}{\sqrt{1+Q_{ab}^2}}\right)} \,\,\, ,
\label{4-14}
\end{equation}
\begin{equation}
\displaystyle A_4 = \arctan{(Q_c)} \,\,\, ,
\label{4-15}
\end{equation}
\begin{equation}
\displaystyle A_5 = \arctan{\left(\frac{1+Q_{ab}^2+Q_c^2}{\Omega}\right)}-
\arctan{\left(\frac{1+Q_c^2}{\Omega}\right)} \,\,\, .
\label{4-16}
\end{equation}
The effects of cutoff are similar as those described at length in
the preceding section for the simpler case of {\it 3D} isotropic
superconductors. In this section we discuss only the modifications
in the limit $\displaystyle T \rightarrow T_c$ where the relevant
parameters can be determined. The ${\cal S}$-functions can be
expanded in the limit of $T_c$ ($\displaystyle \Omega \rightarrow
\infty$), and the leading terms are
\begin{eqnarray}
\displaystyle {\cal S}_{1}^{\hspace{0.02cm}{\it 3D},aniso}
(W_{ab},W_c,\Omega \rightarrow \infty) \approx \frac{4 \sqrt{2}}{3
\pi}& \displaystyle \left[C_1-U_{+}\left(\sqrt{1+W_{ab}^4}
-\frac{1}{2}\, W_{ab}^2\right)\,C_2
+\frac{3}{\sqrt{2}}\,W_c\,C_3+\sqrt{2}\,W_{ab}^{\,3}\,C_4\right.
\label{4-17a}\\
\nonumber &\left. \displaystyle -\frac{1}{2}\,D_1
+\frac{1}{2}\,U_{-}\left(\sqrt{1+W_{ab}^4}+\frac{1}{2}\,W_{ab}^2\right)\,D_2
+\frac{1}{2\sqrt{2}}\,W_c^3\,D_3\right]
\displaystyle \frac{1}{\sqrt{\Omega}} \,\,\, ,
\end{eqnarray}
\begin{eqnarray}
\displaystyle {\cal S}_{2}^{\hspace{0.02cm}{\it 3D},aniso}
(W_{ab},W_c,\Omega \rightarrow \infty) \approx \frac{4 \sqrt{2}}{3
\pi}& \displaystyle \left[C_1-U_{-}\left(\sqrt{1+W_{ab}^4}
+\frac{1}{2}\,W_{ab}^2\right)\,C_2
+\frac{1}{\sqrt{2}}\,W_c^3\,C_3+\frac{3}{\sqrt{2}}\,W_{ab}\,C_4\right.
\label{4-17b}\\
\nonumber &\left. \displaystyle +\frac{1}{2}\,D_1
-\frac{1}{2}\,U_{+}\left(\sqrt{1+W_{ab}^4}-\frac{1}{2}\,W_{ab}^2\right)\,D_2
-\frac{3}{2\sqrt{2}}\,W_c\,D_3\right]
\displaystyle \frac{1}{\sqrt{\Omega}} \,\,\, ,
\end{eqnarray}
where we used the following shorthand notations
\begin{equation}
\displaystyle U_{\pm} = \sqrt{\sqrt{1+W_{ab}^4} \pm W_{ab}^2} \,\,\, ,
\label{4-18}
\end{equation}
\begin{equation}
\displaystyle C_1 =
\arctan{\left(1+\sqrt{2}\,W_c\right)}-\arctan{(1-\sqrt{2}\,W_c)} \,\,\, ,
\label{4-19}
\end{equation}
\begin{equation}
\displaystyle C_2 =
\arctan{\left(\frac{U_{-} + \sqrt{2}\,W_c}{U_{+}}\right)} -
\arctan{\left(\frac{U_{-}-\sqrt{2}\,W_c}{U_{+}}\right)} \,\,\, ,
\label{4-20}
\end{equation}
\begin{equation}
\displaystyle C_3 =
\arctan{\left(W_{ab}^2+W_c^2\right)}-\arctan{\left(W_c^2\right)} \,\,\, ,
\label{4-21}
\end{equation}
\begin{equation}
\displaystyle C_4 = \arctan{\left(\frac{W_c}{W_{ab}}\right)} \,\,\, ,
\label{4-22}
\end{equation}
\begin{equation}
\displaystyle D_1 =
\ln{\left(\frac{1+\sqrt{2}\,W_c+W_c^2}{1-\sqrt{2}\,W_c
+W_c^2}\right)} \,\,\, ,
\label{4-23}
\end{equation}
\begin{equation}
\displaystyle D_2 =
\ln{\left(\frac{\sqrt{1+W_{ab}^4}+\sqrt{2}\,W_c\,U_{-}+W_c^2}
{\sqrt{1+W_{ab}^4}-\sqrt{2}\,W_c\,U_{-}+W_c^2}\right)} \,\,\, ,
\label{4-24}
\end{equation}
\begin{equation}
\displaystyle D_3 =
\ln{\left(\frac{W_c^4[1+(W_{ab}^2+W_c^2)^2]}
{(W_{ab}^2+W_c^2)^2(1+W_c^4)}\right)} \,\,\, .
\label{4-25}
\end{equation}
The cutoff parameters appear in
\begin{equation}
\displaystyle W_{ab} = \sqrt{2}\,\Lambda_{ab}\,
\sqrt{\frac{16}{\pi}\frac{k_B T_c}{\hbar \omega}} \,\,\, ,
\label{4-26}
\end{equation}
\begin{equation}
\displaystyle W_c =
\Lambda_c\,\sqrt{\frac{16}{\pi}\frac{k_B T_c}{\hbar \omega}} \,\,\, .
\label{4-27}
\end{equation}
We note that in the anisotropic case, the $\cal S$-functions
behave also as $1/\sqrt{\Omega}$ when $T \rightarrow T_c$. As
already discussed in the previous section, this implies that
finite nonzero $\sigma_1(T_c)$ and $\sigma_2(T_c)$ can be obtained
only for $z=2$ (relaxational model). Since the available
experimental data in anisotropic high-$T_c$ superconductors
\cite{Anlage:96,Booth:96,Kotzler:97} show finite nonzero
$\sigma_1(T_c)$ and $\sigma_2(T_c)$, we can adopt $z = 2$ in the
remainder of this section.
\par
\begin{figure}
\centerline{\includegraphics[width=0.8\textwidth]{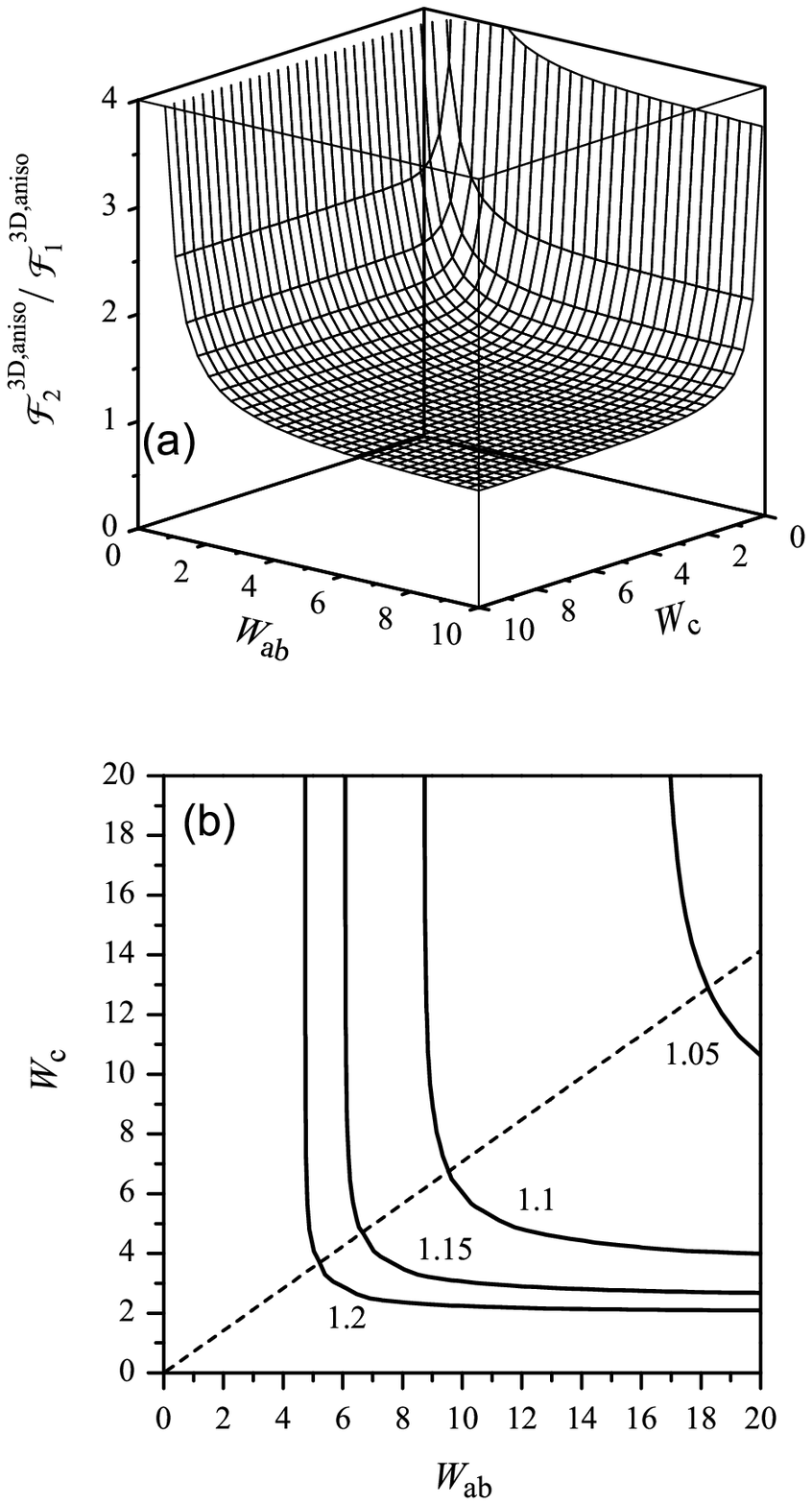}}
\caption{(a) The ratio $\displaystyle {\cal F}_2/{\cal F}_1$ for
{\it 3D} anisotropic case as a function of $W_{ab}$ and $W_c$ (cf.
Eq.~(\ref{4-28})). (b) Selection of curves for fixed ratios of
$\displaystyle {\cal F}_2/{\cal F}_1$ indicated by numbers. The
dashed line marks the condition $\Lambda_{ab} = \Lambda_c$
($W_{ab} = \sqrt{2}\ W_c$).} \label{Fig5}
\end{figure}

In analogy to the {\it 3D} isotropic case described in the
preceding section, one may define the functions
\begin{equation}
\displaystyle {\cal F}_{1,2}^{\hspace{0.02cm}{\it
3D},aniso}(W_{ab},W_c)= \frac{3\,\sqrt{\Omega}}{4\,\sqrt{2}}\,
{\cal S}_{1,2}^{\hspace{0.02cm}{\it 3D},aniso}(W_{ab},W_c,\Omega
\rightarrow \infty) \label{4-28}
\end{equation}
so that the conductivities at $T_c$ are given by
\begin{equation}
{\sigma}_{1,2}^{\hspace{0.02cm}{\it
3D},aniso}(\omega,T_c,\Lambda_{ab},\Lambda_c)=
\frac{e^2}{6\hbar\xi_{0c}}\sqrt{\frac{2}{\pi}\frac{k_B T_c}{\hbar
\omega}}\, {\cal F}_{1,2}^{\hspace{0.02cm}{\it
3D},aniso}(W_{ab},W_c) \,\,\, . \label{4-29}
\end{equation}
The ratio of experimental values $\displaystyle \sigma_2
/\sigma_1$ at $T_c$ does not define uniquely the cutoff parameters
$\Lambda_{ab}$ and $\Lambda_c$. It puts, however, a constraint on
their choice. Fig.~\ref{Fig5}(a) shows the plot of $\displaystyle
{\cal F}_2 /{\cal F}_1$ given by Eq.~(\ref{4-28}) as a function of
two variables, $W_{ab}$ and $W_c$. It is evident that a fixed
value of $\displaystyle {\cal F}_2 /{\cal F}_1$ defines a simple
curve of the possible choices of ($W_{ab}$,$W_c$).
Fig.~\ref{Fig5}(b) shows a selection of such curves for
$\displaystyle {\cal F}_2 /{\cal F}_1 = $1.05, 1.1, 1.15, and 1.2.
The dashed line marks the condition $\displaystyle \Lambda_{ab} =
\Lambda_c$ ($\displaystyle W_{ab} = \sqrt{2}\, W_c$).
Experimentally, one has to probe the possible choices for
($W_{ab}$, $W_c$) and look at the fits of the theoretical curves
to the experimental data at $\displaystyle T > T_c$. The parameter
$\xi_{0c}$ in Eq.~(\ref{4-5}) can be obtained once the choice for
($W_{ab}$, $W_c$) is made.
\par
Note that in practical applications of the above theory one needs
measurements where the microwave current flows only in the ab-plane.
Particularly suitable for this purpose are the measurements in which
the superconducting sample is placed in the antinode of the
microwave electric field $E_{\omega}$ in the cavity.
\cite{Peligrad:98,Peligrad:01}
\section{{\it 2D} FLUCTUATIONS}
\label{Sect4}
Superconducting transition does not occur in a strictly {\it 2D}
system. However, if the sample is a very thin film so that its
thickness is much smaller than the correlation length, the
fluctuations will be restricted within the film thickness $d$ in
one direction, and develop freely only in the plane of the film.
Using the formalism described in the preceding sections, we find
that the fluctuation conductivity is given by
\begin{equation}
\displaystyle {\sigma_1}^{\hspace{-0.1cm}{\it
2D}}(\omega,T,\Lambda)= \frac{e^2}{4 \hbar
d}\left(\frac{\xi(T)}{\xi_0}\right)^2 \sum_{q_{n}}\int_0^{Q}
\frac{q^3} {(1+q^2+q_{n}^2)[\Omega^2+(1+q^2+q_{n}^2)^2]}\,dq
\,\,\, , \label{5-1}
\end{equation}
\begin{equation}
\displaystyle {\sigma_2}^{\hspace{-0.1cm}{\it
2D}}(\omega,T,\Lambda)= \frac{e^2}{4 \hbar d}
\left(\frac{\xi(T)}{\xi_0}\right)^2
\sum_{q_{n}}\int_0^{Q}\frac{\Omega\,q^3}
{(1+q^2+q_{n}^2)^2[\Omega^2+(1+q^2+q_{n}^2)^2]}\,dq \,\,\, ,
\label{5-2}
\end{equation}
\begin{equation}
\displaystyle {\widetilde\sigma}^{{\it 2D}}(\omega,T,\Lambda)=
\frac{e^2}{16 \hbar d}\left(\frac{\xi(T)}{\xi_0}\right)^2
\sum_{q_{n}}\left[{{\cal S}_1}^{\hspace{-0.1cm}{\it
2D}}(\Omega,Q,q_{n})+ i\, {{\cal S}_2}^{\hspace{-0.1cm}{\it
2D}}(\Omega,Q,q_{n})\right] \,\,\, , \label{5-3}
\end{equation}
where
\begin{equation}
\displaystyle q_{n} = n\,\pi\left(\frac{\xi_0}{d}\right)
\left(\frac{\xi (T)}{\xi_0}\right) \,\,\, ,
\label{5-4}
\end{equation}
and $\displaystyle Q = \sqrt{2}\ \Lambda\ (\xi (T)/\xi_0)$. The
prefactor is the Aslamazov-Larkin result for the {\it 2D} case
with no cutoff, and the Gaussian form  $1/\epsilon$ replaced by
the more general expression $\displaystyle (\xi(T)/ \xi_0)^{2}$.
\par
The ${\cal S}$-functions are given by
\begin{eqnarray}
\displaystyle {\cal S}_1^{\hspace{0.05cm}{\it 2D}}(\Omega,Q,q_{n})
= \frac{1}{\Omega}& \displaystyle
\left[2\,\arctan{\left(\frac{1+Q^2+q_{n}^2}{\Omega}\right)}
-2\,\arctan{\left(\frac{1+q_{n}^2}{\Omega}\right)}\right.
\label{5-5}\\
\nonumber &\left. \displaystyle
+\left(\frac{1+q_{n}^2}{\Omega}\right)
\ln{\left(\frac{(1+q_{n}^2)^2[\Omega^2+(1+Q^2+q_{n}^2)^2]}
{(1+Q^2+q_{n}^2)^2[\Omega^2+(1+q_{n}^2)^2]}\right)}\right] \,\,\, ,
\end{eqnarray}
\begin{eqnarray}
\displaystyle {\cal S}_2^{\hspace{0.05cm}{\it 2D}}(\Omega,Q,q_{n})
= \frac{1}{\Omega}& \displaystyle
\left[\frac{2(1+q_{n}^2)}{\Omega}\,
\left[\arctan{\left(\frac{1+Q^2+q_{n}^2}{\Omega}\right)} -
\arctan{\left(\frac{1+q_{n}^2}{\Omega}\right)}\right]\right.
\label{5-6}\\
\nonumber &\left. \displaystyle
-\ln{\left(\frac{(1+q_{n}^2)^2[\Omega^2+(1+Q^2+q_{n}^2)^2]}
{(1+Q^2+q_{n}^2)^2[\Omega^2+(1+q_{n}^2)^2]}\right)}
-2\,\frac{Q^2}{1+Q^2+q_{n}^2}\right] \,\,\, .
\end{eqnarray}
\par
The summation over $q_{n}$ in Eq.~(\ref{5-3}) has to carried out
until the factor $n \pi (\xi_0/d)$ reaches some cutoff value
$\Lambda$, which is of the order of unity. If the film thickness
is large ($d \gg \xi_0$), one has to sum up to a high $n$-value.
In such cases, the summation is well approximated by an
integration, and one retrieves the {\it 3D} case of the preceding
section. The {\it 2D} character is better displayed when the film
thickness is comparable to $\xi_0$. Then, only a few terms have to
be taken into account. In the extreme case of $d < \xi_0$, only
the $n = 0$ term is found below the cutoff limit.
\par
The zero frequency limit ($\Omega \rightarrow 0$) yields
\begin{equation}
\displaystyle {\sigma_{dc}}^{\hspace{-0.25cm}{\it 2D}}(T,\Lambda)=
\frac{e^2}{16 \hbar d}\left(\frac{\xi(T)}{\xi_0}\right)^2
\sum_{q_{n}}\frac{Q^4}{(1+Q^2+q_{n}^2)^2 (1+q_{n}^2)} \,\,\, .
\label{5-7}
\end{equation}
The $n = 0$ term yields the previous result of
Hopfeng{\"a}rtner et al. \cite{Hopfengartner:91} and Gauzzi et al.
\cite{Gauzzi:95}
\par
In the limit of $T_c$ ($\Omega \rightarrow \infty$) one obtains
\begin{equation}
\displaystyle {\sigma_1}^{\hspace{-0.1cm}{\it 2D}}(T_c)=
\frac{e^2}{\pi \hbar d}\frac{k_B T_c}{\hbar \omega}
\sum_{n}\left[2\,\arctan{(W^2+W_{n}^2)}-2\,\arctan{(W_{n}^2)}
+W_{n}^2\,\ln{\frac{W_{n}^4\left[1+(W^2+W_{n}^2)^2 \right]}
{(1+W_{n}^4)(W^2+W_{n}^2)^2}}\right] \,\,\, , \label{5-8}
\end{equation}
\begin{equation}
\displaystyle {\sigma_2}^{\hspace{-0.1cm}{\it 2D}}(T_c)=
\frac{e^2}{\pi \hbar d}\frac{k_B T_c}{\hbar \omega}
\sum_{n}\left[2\,W_{n}^2\,\arctan{(W^2+W_{n}^2)}
-2\,W_{n}^2\,\arctan{(W_{n}^2)}
-\ln{\frac{W_{n}^4\left[1+(W^2+W_{n}^2)^2 \right]}
{(1+W_{n}^4)(W^2+W_{n}^2)^2}}\right] \,\,\, , \label{5-9}
\end{equation}
where we used the notation
\begin{equation}
\displaystyle W = \sqrt{2}\,\Lambda
\sqrt{\frac{16}{\pi}\frac{k_B T_c}{\hbar \omega}} \,\,\, ,
\label{5-10}
\end{equation}
\begin{equation}
\displaystyle W_{n} = n\,\pi\left(\frac{\xi_0}{d}\right)
\sqrt{\frac{16}{\pi}\frac{k_B T_c}{\hbar \omega}} \,\,\, .
\label{5-11}
\end{equation}
One may observe that for $n = 0$ the real part of the conductivity
is finite, but the imaginary part diverges. It is due to the
logarithmic term in Eq.~(\ref{5-9}). This is an unphysical result.
It may indicate that the $n = 0$ term is not physically
acceptable, or that the {\it 2D} model should not be applied
exactly at $T_c$.

\section{COMPARISON WITH EXPERIMENT}
\label{Sect5}

\begin{figure}
\centerline{\includegraphics[width=0.8\textwidth]{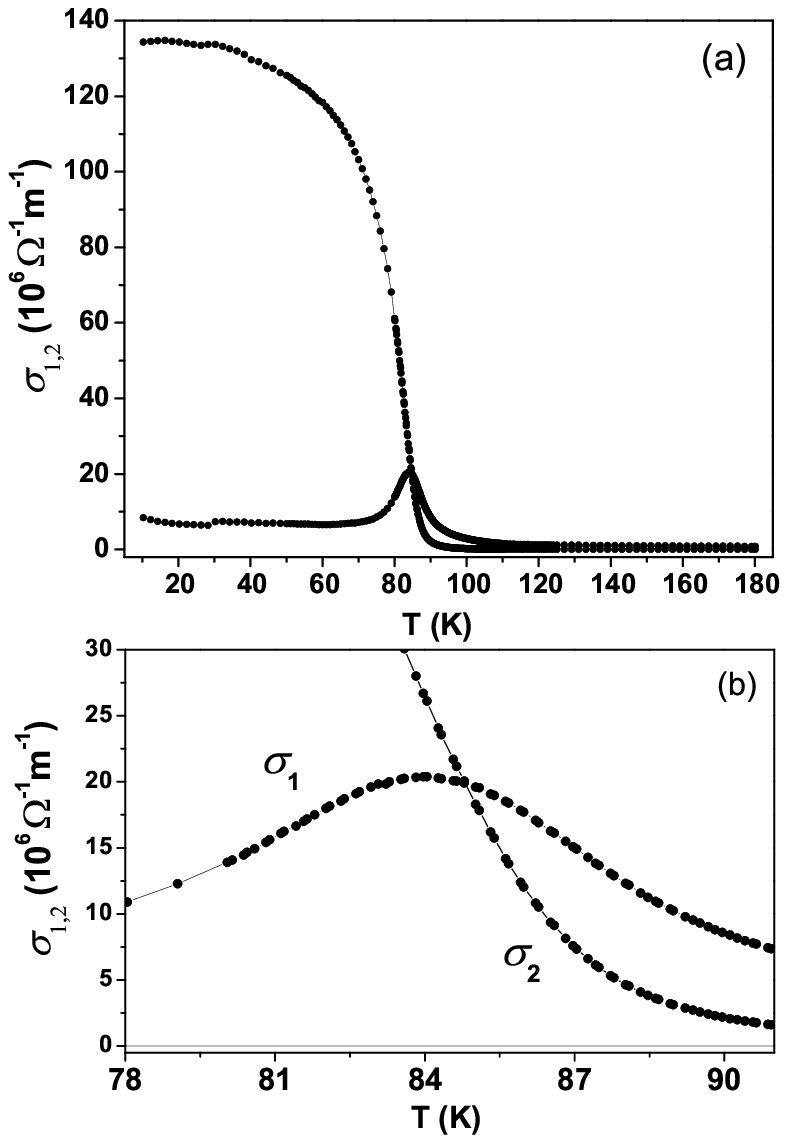}}
\caption{(a)
Experimental complex conductivity in
$Bi_2Sr_2Ca_2Cu_3O_{10-\delta}$ thin film at 9.5 GHz. (b) Enlarged
section near $T_c$.} \label{Fig6}
\end{figure}

The relevance of the theoretical expressions derived in the
preceding sections can be demonstrated by comparison of the
calculated and experimentally measured $ac$ fluctuation
conductivity. As an example we present here an analysis of the
data in $Bi_2Sr_2Ca_2Cu_3O_{10-\delta}$ thin film. The
experimental results of the complex conductivity measured at 9.5
GHz are shown in Fig.~\ref{Fig6}(a). The main features are the
same as reported previously in single crystals of high-$T_c$
superconductors.\cite{Anlage:96,Waldram:99} We have to note that
in our measurement the thin film was positioned in the center of
an elliptical microwave cavity resonating in $_{\rm e}$TE$_{111}$
mode, and oriented in such a way that the electric field
$E_{\omega}$ was in the {\it ab}-plane. Thus the in-plane
conductivity was measured and the application of the theoretical
expressions of the preceding sections is appropriate. Other
experimental details have been reported
previously.\cite{Peligrad:01,Nebendahl:01,Peligrad:02} In this
section we are interested in the fluctuation conductivity near
$T_c$ which is shown on an enlarged scale in Fig.~\ref{Fig6}(b).
The real part of the conductivity has a maximum when the coherence
length diverges. Since the critical temperature of a phase
transition is characterized by the divergence of the coherence
length, we use the maximum of $\sigma_1$ in Fig.~\ref{Fig6}(b) to
determine $T_c = 84.04$ K. One can also observe in
Fig.~\ref{Fig6}(b) that the imaginary part of the conductivity
crosses the real part at a temperature slightly above $T_c$. This
is a direct experimental evidence of the short wavelength cutoff
as discussed in Section~\ref{Sect2}. The experimental values of
$\sigma_1$ and $\sigma_2$ at $T_c$ can be used in the evaluation
of the parameters which enter the theoretical expressions of the
preceding sections.
\par
\begin{figure}
\centerline{\includegraphics[width=0.8\textwidth]{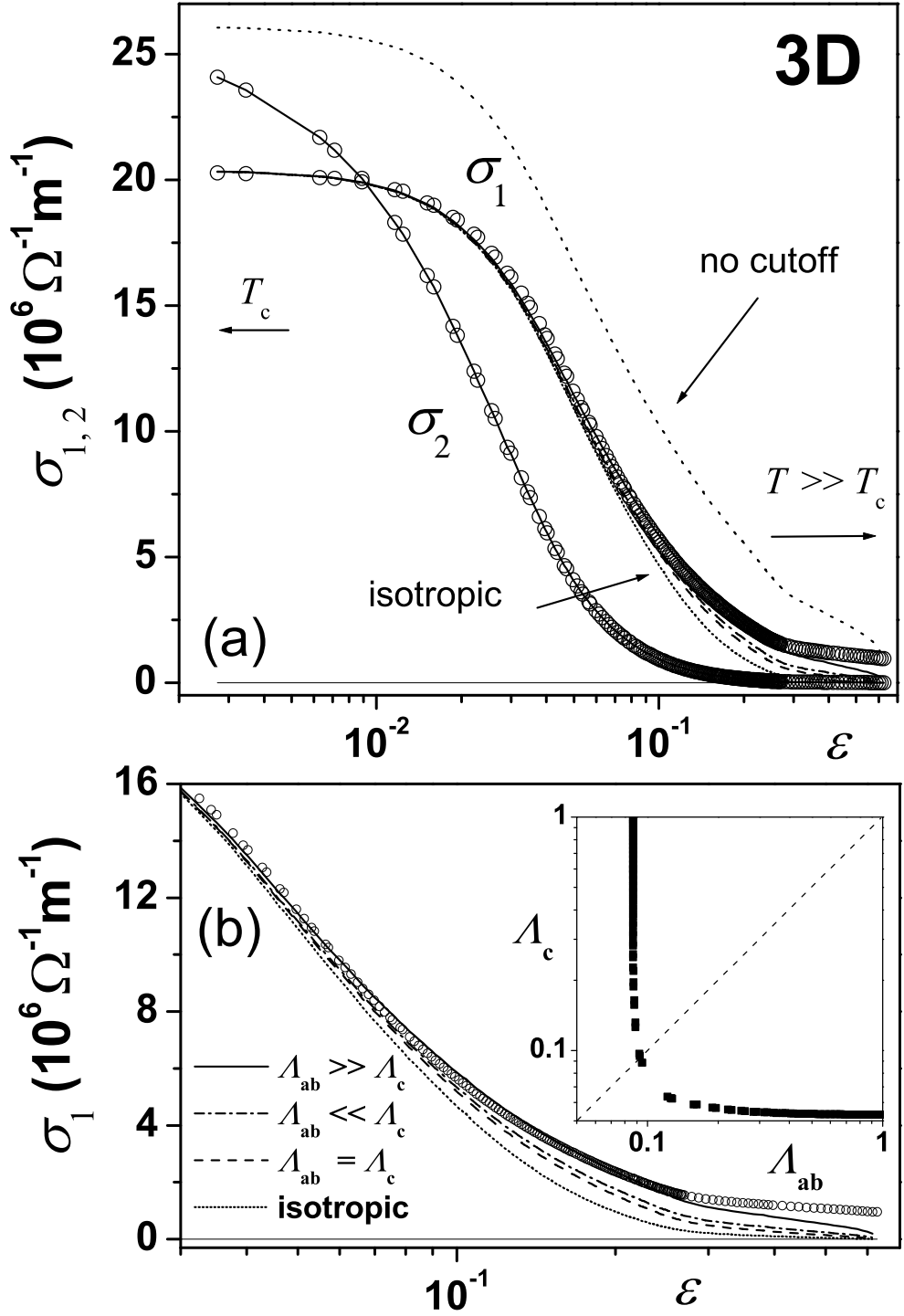}}
\caption{(a)
Experimental data (symbols) of the complex conductivity from
Fig.~\ref{Fig6} above $T_c$ plotted versus the reduced temperature
$\displaystyle \epsilon=\left(\ln{T/T_c}\right)$. Various lines
are the conductivities calculated in the {\it 3D} cases as
described in the text. (b) Enlarged view of the high temperature
part of the same curves as in (a). The constraining curve for the
choices of the cutoff parameters $\Lambda_{ab}$ and $\Lambda_c$
resulting from the experimental ratio
$\sigma_2(T_c)/\sigma_1(T_c)= 1.28$ at $T_c$ is shown in the
inset.} \label{Fig7}
\end{figure}

Fig.~\ref{Fig7}(a) shows the experimental data above $T_c$ plotted
against the reduced temperature $\displaystyle \epsilon =
\ln{\left(T/T_c\right)}$. We can analyze this data first by the
theoretical expressions which have no cutoff on the fluctuation
wavevector. If one sets $\Lambda_{ab} = \Lambda_c = \infty$ for
the {\it 3D} anisotropic case, it is straightforward to evaluate
$\xi_{0c}$ using Eq.~(\ref{4-29}) and the experimental value of
$\sigma_2$ at $T_c$. We have obtained $\xi_{0c} = 0.016$ nm in
$Bi_2Sr_2Ca_2Cu_3O_{10-\delta}$ thin film. Once this parameter is
determined, the fluctuation conductivity at all temperatures above
$T_c$ follows from Eq.~(\ref{4-5}). The real and imaginary parts
of the $\it ac$ fluctuation conductivity have to be mutually
consistent. This can be exploited in the data analysis. We insert
the experimental values of $\sigma_2$ into the imaginary part of
Eq.~(\ref{4-5}), and  solve numerically for $\xi(T)/\xi_0$. Then
we exploit these same values of the reduced correlation length in
the real part of Eq.~(\ref{4-5}). The calculated $\sigma_1$ is
shown by the dotted line in Fig.~\ref{Fig7}(a). The calculated
line lies far from the experimental points. Note in particular
that the calculated $\sigma_1$ meets $\sigma_2$ at $T_c$ when no
cutoff is included (cf. Section~\ref{Sect2}). Besides, the shape
of the calculated $\sigma_1$ differs from that of the experimental
one. We may conclude that with no cutoff on the fluctuation
wavevector the theoretical expression does not describe properly
the experimental fluctuation conductivity.
\par
A finite cutoff on the fluctuation wavevector improves greatly the
agreement of the theory and experiment. From Fig.~\ref{Fig6}(b) we
can evaluate the ratio $\sigma_2(T_c)/\sigma_1(T_c) = 1.28$ at
$T_c$. This value yields a constraint on the choices of
$\Lambda_{ab}$ and $\Lambda_c$ as described in Section~\ref{Sect3}
and Fig.~\ref{Fig5}. The actual choices are presented in the inset
of Fig.~\ref{Fig7}(b). For a given choice ($\Lambda_{ab}$,
$\Lambda_c$) from this constraining line, one has to determine
first $\xi_{0c}$ using Eq.~(\ref{4-29}) and the experimental value
of $\sigma_2$ at $T_c$. Then, the temperature dependence of the
reduced correlation length $\xi(T)/\xi_0$ is evaluated numerically
from the imaginary part of Eq.~(\ref{4-5}) with the selected pair
($\Lambda_{ab}$, $\Lambda_c$) and the experimental values of
$\sigma_2$. The obtained values of $\xi(T)/\xi_0$ are finally used
to calculate $\sigma_1$ from the real part of Eq.~(\ref{4-5}). The
results vary with the possible choices of the pairs
($\Lambda_{ab}$, $\Lambda_c$) from inset of Fig.~\ref{Fig7}(b).
The best fit of the calculated $\sigma_1$ to the experimental
points is shown by the solid line in Fig.~\ref{Fig7}(a). It is
obtained with the choice ( $\Lambda_{ab} = 0.71$, $\Lambda_c =
0.05$ ). It is physically reasonable. With $\Lambda_{ab}$ being of
the order of unity, the minimum wavelength of the superconducting
fluctuations in the {\it ab}-plane is given by $2 \pi \xi_{0ab}$,
which is much larger than the atomic size and could be accepted as
a mesoscopic quantity. In contrast, the value of $2 \pi \xi_{0c}$
is below the atomic size and, hence, could not be physically
accepted for the lower limit of the fluctuation wavelength along
the c-axis. Therefore one needs a value of $\Lambda_c << 1$ so
that the minimum wavelength of the superconducting fluctuations $2
\pi \xi_{0c}/\Lambda_c$ along the c-axis becomes also an
acceptable mesoscopic quantity.
\par
We have tested also a number of other choices of the cutoff
parameters. By shifting the choice of the parameters along the
constraining curve in the inset of Fig.~\ref{Fig7}b, one degrades
the fit of the calculated $\sigma_1$ to the experimental points.
The dashed line in Fig.~\ref{Fig7}(a) is the calculated $\sigma_1$
using the choice of equal cutoff parameters ( $\Lambda_{ab} =
\Lambda_c = 0.09$ ) permitted by the constraint in the inset of
Fig.~\ref{Fig7}(b). The fit to the experimental points is seen to
be much worse than that of the full line. For the sake of
completeness, we show also the result for a choice of the cutoff
parameters on the other branch of the constraining curve in the
inset of Fig.~\ref{Fig7}(b). If one takes ($\Lambda_{ab} = 0.08$,
$\Lambda_c = 0.71$ ) the calculated $\sigma_1$ is as shown by the
dashed-dotted line in Fig.~\ref{Fig7}(a). The fit is
unsatisfactory. Moreover, this choice has to be refuted on the
grounds of physically unacceptable minimum wavelength of the
fluctuations along the c-axis. Also shown in Fig.~\ref{Fig7}(a) by
the short-dotted line is the result obtained by the isotropic {\it
3D} expression in Eq.~(\ref{3-8}). In this case the parameters
$\xi_0 = 0.016$ nm and $\Lambda = 0.08$ are obtained
straightforwardly from Eq.~(\ref{3-19}) and the experimental
values of $\sigma_1$ and $\sigma_2$ at $T_c$. The fit in
Fig.~\ref{Fig7}(a) is obviously not good. It is also seen that the
expressions for the anisotropic {\it 3D} case always yield curves
which are different from that of the isotropic {\it 3D} case. The
complexity of the anisotropic {\it 3D} expressions elaborated in
Section~\ref{Sect3} is not futile. Indeed, we find that these
expressions must be used when analyzing an anisotropic
superconductor. Fig.~\ref{Fig7}(b) shows an enlarged view of the
high temperature part where the same curves as in
Fig.~\ref{Fig7}(a) are better distinguished.
\par
\begin{figure}
\centerline{\includegraphics[width=0.8\textwidth]{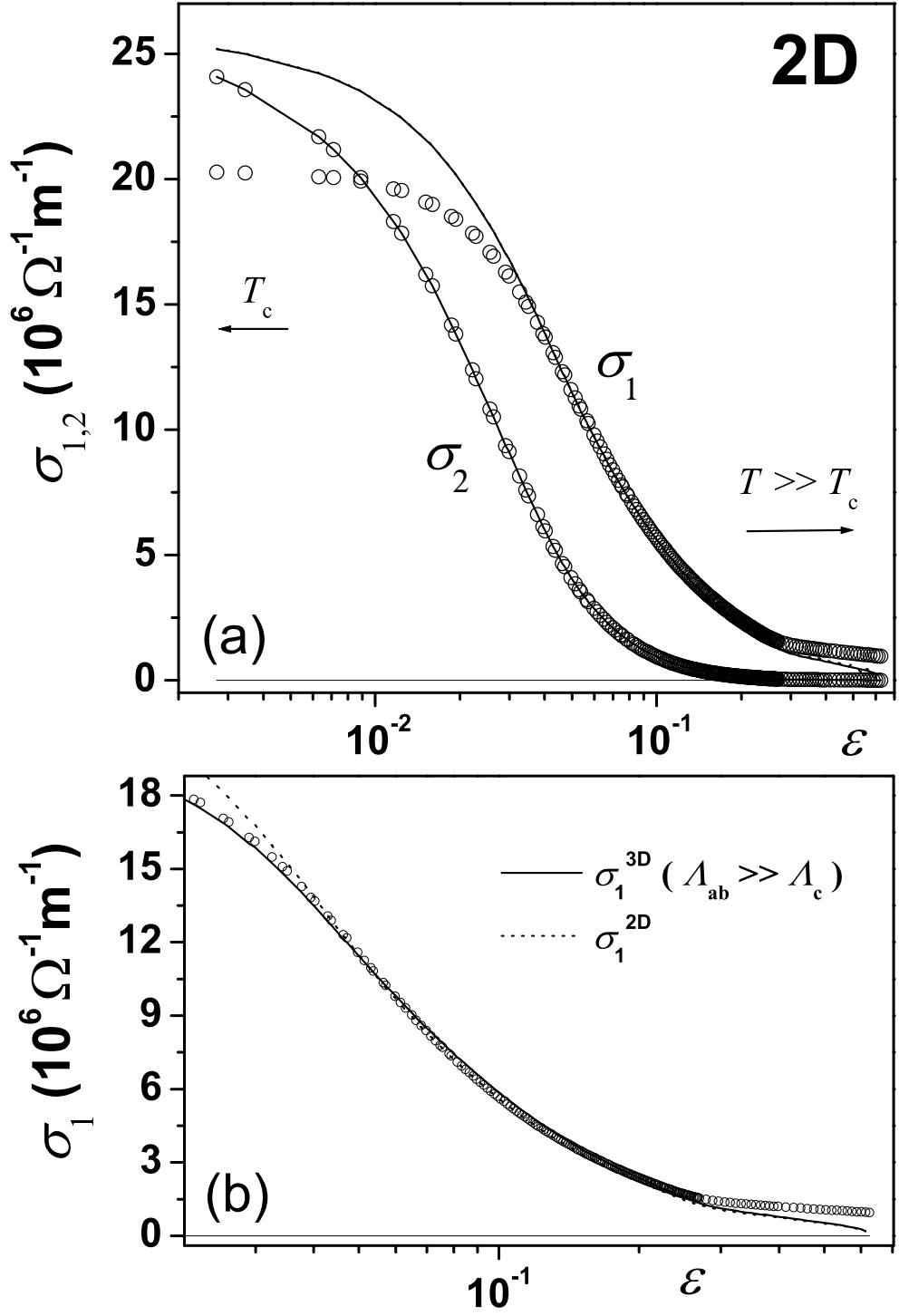}}
\caption{(a) Experimental data (symbols) as in Fig.~\ref{Fig7}(a)
with the calculated {\it 2D} curve as described in the text. \\
(b) The superposition of the calculated {\it 2D} and {\it 3D}
curves.} \label{Fig8}
\end{figure}

We have analyzed also the {\it 2D} expressions of
Section~\ref{Sect4}. Fig.~\ref{Fig8}(a) shows again the same
experimental data as in Fig.~\ref{Fig7}(a), but fitted with $n=0$
term of the {\it 2D} expansion in Eq.~(\ref{5-3}). The parameter
$d$ has been chosen so as to optimize the fit to the experimental
values of $\sigma_1$. The resulting curve in Fig.~\ref{Fig8}(a)
was obtained with $d = 1.2$ nm. The {\it 2D} results are not so
sensitive to the fluctuation wavevector cutoff as those of the
{\it 3D} case. The curves obtained with no cutoff ($\Lambda =
\infty$) and with $\Lambda = 0.7$ are practically
indistinguishable in Fig.~\ref{Fig8}(a). One may conclude that
closer to $T_c$ the BSCCO superconductor clearly does not behave
as a {\it 2D} system. However, at higher temperatures both, {\it
2D} and {\it 3D} expressions yield almost equally good fits to the
experimental values, as seen in Fig.~\ref{Fig8}(b). Thus the
dimensionality of the fluctuations at higher temperatures cannot
be resolved from the ${\it ac}$ fluctuation conductivity.
\par
Finally, we remark that the above analysis could explain very well
the experimental ${\it ac}$ fluctuation conductivity above $T_c$
in the BSCCO thin film using the Aslamazov-Larkin type expressions
with wavevector cutoff as deduced in the preceding sections of
this paper. The other contributions such as Maki-Thomson and one
electron density of states, mentioned in Section~\ref{Sect2}, are
not necessary over most of the temperature range covered in the
experiment. This is in accord with the recent microscopic
calculation\cite{Livanov:00} proving that these contributions may
cancel in the ultraclean case of nonlocal electrodynamics.
However, they may play a role at high enough temperatures where
the above calculated curves depart from the experimental data.
Their analysis is beyond the scope of the present paper.
\section{CONCLUSIONS}
\label{Sect6}
We have presented full analytical expressions for the $\it{ac}$
fluctuation conductivity in {\it 3D} isotropic, {\it 3D}
anisotropic, and {\it 2D} superconductors. The effects of the
short wavelength cutoff in the fluctuation spectrum on the
$\it{dc}$ and $\it{ac}$ conductivities were discussed in detail.
The short wavelength cutoff brings about a breakdown of the
scaling property in frequency and temperature. It also has a
small, but experimentally very important effect on $\sigma_1$. Due
to a finite cutoff parameter $\Lambda$, the value of $\sigma_1$ at
$T_c$ is lower than that of $\sigma_2$. In the simpler {\it 3D}
isotropic case, this observation can be used to determine
$\Lambda$ directly from the experimental data at $T_c$. In the
{\it 3D} anisotropic case, one obtains a constraint on the choices
of ($\Lambda_{ab}$, $\Lambda_c$).
\par
The useful feature of $\it{ac}$ fluctuation conductivity
measurements is that $T_c$ can be determined directly from the
experimental data. Moreover, the expressions derived in this paper
enable the determination of $\xi_0$ ({\it 3D} isotropic), or
$\xi_{c0}$ ({\it 3D} anisotropic) from the experimental values
$\sigma_{1,2} (T_c)$. Thus, we establish that the analysis of
$\it{ac}$ fluctuation conductivity requires no free fit parameters
in the {\it 3D} isotropic case, and only one ($\Lambda_{ab}$,
$\Lambda_c$) in the {\it 3D} anisotropic case.
\par
We have shown that the anisotropic {\it 3D} expressions with an
appropriate cutoff of the fluctuation wavevector can account for
the experimental fluctuation conductivity in a BSCCO thin film
within a large temperature range above $T_c$. The {\it 2D}
expression is less sensitive to the cutoff and was found to match
the experimental data only at higher temperatures.
\acknowledgments D.-N.~Peligrad and M.~Mehring acknowledge support
by the Deutsche Forschungsgemeinschaft~(DFG) project
Nr.~Me362/14-2. A.~Dul\v{c}i\'{c} acknowledges support from the
Croatian Ministry of Science, and DLR~Stiftung (project
Nr.~KRO--004--98).

\bibliographystyle{apsrev}

\newpage

\end{document}